\documentclass[a4paper,11pt]{article}
\pdfoutput=1 

\usepackage{jcappub} 

\usepackage[T1]{fontenc} 

\usepackage{graphicx}
 \usepackage{dcolumn}
\usepackage{bm}
\usepackage[usenames]{color}
\usepackage[normalem]{ulem}
\usepackage{hyperref}
\usepackage{braket} 
\usepackage{amsmath}
\usepackage{cancel}

\usepackage{bbold}

\begin{document}
\setcounter{equation}{0}

\newcommand{\fref}[1]{Fig.~\ref{fig:#1}} 
\newcommand{\eref}[1]{Eq.~\eqref{eq:#1}} 
\newcommand{\erefn}[1]{ (\ref{eq:#1})}
\newcommand{\erefs}[2]{Eqs.~(\ref{eq:#1}) - (\ref{eq:#2}) } 
\newcommand{\aref}[1]{Appendix~\ref{app:#1}}
\newcommand{\sref}[1]{Section~\ref{sec:#1}}
\newcommand{\cref}[1]{Chapter~\ref{ch:.#1}}
\newcommand{\tref}[1]{Table~\ref{tab:#1}}

\newcommand{\nn}{\nonumber \\}  
\newcommand{\nnl}{\nonumber \\}  
\newcommand{\nl}{& \nonumber \\ &}
\newcommand{\bnl}{\right .  \nonumber \\  \left .}
\newcommand{\dbnl}{\right .\right . & \nonumber \\ & \left .\left .}

\newcommand{\beq}{\begin{equation}} 
\newcommand{\eeq}{\end{equation}} 
\newcommand{\ba}{\begin{array}}  
\newcommand{\ea}{\end{array}} 
\newcommand{\bea}{\begin{eqnarray}}  
\newcommand{\eea}{\end{eqnarray} }  
\newcommand{\be}{\begin{eqnarray}}  
\newcommand{\ee}{\end{eqnarray} }  
\newcommand{\bal}{\begin{align}}
\newcommand{\eal}{\end{align}}   
\newcommand{\bi}{\begin{itemize}}  
\newcommand{\ei}{\end{itemize}}  
\newcommand{\ben}{\begin{enumerate}}  
\newcommand{\een}{\end{enumerate}}  
\newcommand{\bc}{\begin{center}}
\newcommand{\ec}{\end{center}} 
\newcommand{\bt}{\begin{table}}
\newcommand{\et}{\end{table}}  
\newcommand{\btb}{\begin{tabular}}
\newcommand{\etb}{\end{tabular}}  
\newcommand{\bvec}{\left ( \ba{c}}
\newcommand{\evec}{\ea \right )}

\newcommand{\cO}{{\mathcal O}} 
\newcommand{\co}{{\mathcal O}} 
\newcommand{\cL}{{\mathcal L}} 
\newcommand{\cl}{{\mathcal L}} 
\newcommand{\cM}{{\mathcal M}}

\newcommand{\const}{\mathrm{const}}

\newcommand{\ev}{ \mathrm{eV}}
\newcommand{\kev}{\mathrm{keV}}
\newcommand{\mev}{\mathrm{MeV}}
\newcommand{\gev}{\mathrm{GeV}}
\newcommand{\tev}{\mathrm{TeV}}

\newcommand{\mpl}{M_{\mathrm Pl}}

\def\mgut{\, M_{\rm GUT}}
\def\tgut{\, t_{\rm GUT}}
\def\mpl{\, M_{\rm Pl}}
\def\mkk{\, M_{\rm KK}}
\newcommand{\msusy}{M_{\rm soft}}

\newcommand{\dslash}[1]{#1 \! \! \! {\bf /}}
\newcommand{\ddslash}[1]{#1 \! \! \! \!  {\bf /}}

\def\ads{AdS$_5$\,}
\def\adse{AdS$_5$}
\def\intdk{\int {d^4 k \over (2 \pi)^4}} 

\def\ra{\rangle}
\def\la{\langle}  

\def\sgn{{\rm sgn}}
\def\pa{\partial}  
\newcommand{\dlr}{\overleftrightarrow{\partial}}
\newcommand{\Dlr}{\overleftrightarrow{D}}
\newcommand{\re}{{\mathrm{Re}} \,}
\newcommand{\im}{{\mathrm{Im}} \,}
\newcommand{\tr}{\mathrm T \mathrm r}  

\newcommand{\Ra}{\Rightarrow}
\newcommand{\lra}{\leftrightarrow}
\newcommand{\llra}{\longleftrightarrow}

\newcommand\simlt{\stackrel{<}{{}_\sim}}
\newcommand\simgt{\stackrel{>}{{}_\sim}}   
\newcommand{\zt}{$\mathbb Z_2$ }

\newcommand{\ha}{{\hat a}}
\newcommand{\hab}{{\hat b}}
\newcommand{\hac}{{\hat c}} 

\newcommand{\ti}{\tilde}  
\def\hc{{\rm h.c.}} 
\def\ov{\overline}  
  

\newcommand{\eps}{\epsilon}
\newcommand{\eS}{\epsilon_S}
\newcommand{\eT}{\epsilon_T}
\newcommand{\eP}{\epsilon_P}
\newcommand{\eL}{\epsilon_L}
\newcommand{\eR}{\epsilon_R}
\newcommand{\teps}{{\tilde{\epsilon}}}
\newcommand{\teS}{{\tilde{\epsilon}_S}}
\newcommand{\teT}{{\tilde{\epsilon}_T}}
\newcommand{\teP}{{\tilde{\epsilon}_P}}
\newcommand{\teL}{{\tilde{\epsilon}_L}}
\newcommand{\teR}{{\tilde{\epsilon}_R}}
\newcommand{\eLc}{{\epsilon_L^{(c)}}}
\newcommand{\eLv}{{\epsilon_L^{(v)}}}
\newcommand{\eSP}{\epsilon_{S,P}}
\newcommand{\teSP}{{\tilde{\epsilon}_{S,P}}}

\newcommand{\lz}{\lambda_z}
\newcommand{\dgz}{\delta g_{1,z}}
\newcommand{\dkg}{\delta \kappa_\gamma}

\def\cog{\color{OliveGreen}}
\def\cor{\color{Red}}
\def\copu{\color{purple}}
\def\coro{\color{RedOrange}}
\def\coma{\color{Maroon}}
\def\cob{\color{Blue}}
\def\cobr{\color{Brown}}
\def\cobl{\color{Black}}
\def\cost{\color{WildStrawberry}}

\newcommand{\tl}{{\tilde{\lambda}}}
\newcommand{\dll}{{\frac{\delta\lambda}{\lambda}}}

\title{\boldmath
Flavor Triangle of the Diffuse Supernova Neutrino Background
}

\author{Zahra~Tabrizi,}
\author{Shunsaku Horiuchi}

\affiliation{Center for Neutrino Physics, Department of Physics, Virginia Tech, Blacksburg, VA 24061, USA}

\emailAdd{ztabrizi@vt.edu}
\emailAdd{horiuchi@vt.edu}

\abstract{Although Galactic core-collapse supernovae (SNe) only happen a few times per century, every hour a vast number of explosions happen in the whole universe, emitting energy in the form of neutrinos, resulting in the diffuse supernova neutrino background (DSNB). The DSNB  has not yet been detected, but Super-Kamiokande doped with gadolinium is expected to yield the first statistically significant observation within the next  several years. 
Since the neutrinos produced at the core collapse undergo mixing during their propagation to Earth, the flavor content at detection is a test of oscillation physics.
In this paper, we estimate the expected DSNB data at the DUNE, Hyper-K and JUNO experiments which when combined are sensitive to all different neutrino flavors. 
We determine how well the flavor content of the DSNB will be reconstructed in the future, for a Mikheyev-Smirnov-Wolfenstein (MSW) scenario as well as a neutrino decay scenario. A large fraction of the flavor space will be excluded, but the heavy-lepton neutrino flux remains a challenge.
}

\maketitle

\flushbottom

\section{Introduction}

When a massive star enters the end of its life, a core collapse will happen. During this process, neutrino and anti-neutrino of all flavors are produced at the core of the newly formed proto-neutron star. These neutrinos carry away the bulk of the tremendous amount of energy liberated in the collapse ($\sim 10^{53}$~ergs), and are emitted over a period of several seconds (for reviews, see, e.g., Refs.~\cite{Langanke:2002ab,Mezzacappa:2005ju,Kotake:2005zn,Woosley:2006ie,Foglizzo:2015dma,Janka:2017vlw,Burrows:2020qrp}). Eventually, the collapsed core settles to a neutron star (NS) or a black hole (BH). In the core-collapsed core, the heavy lepton neutrinos attain similar energy spectra. Therefore, the emitted neutrinos are often described by three neutrino fluxes, $\nu_e$, $\bar\nu_e$ and $\nu_x$, where the latter collectively denotes  $\nu_\mu$, $\bar\nu_\mu$, $\nu_\tau$ and $\bar\nu_\tau$ (but, see \cite{Nagakura:2020gls} for a possible acceleration mechanism which distinguishes $\nu_\mu$'s and $\nu_\tau$'s).

After production, the neutrinos will flavor mix on their way to Earth, through ($i$) collective flavor oscillation, which happens within a few hundred kilometers from the core due to $\nu-\nu$ coherent scattering \cite{Duan:2006an,Hannestad:2006nj,Fogli:2007bk,Dasgupta:2009mg,EstebanPretel:2007ec,Dasgupta:2008my,Dasgupta:2011jf,Chakraborty:2015tfa,Dasgupta:2016dbv,Izaguirre:2016gsx,Capozzi:2018clo}, ($ii$) through the coherent forward scattering on electrons of the stellar matter, described by the Mikheyev-Smirnov-Wolfenstein (MSW) mechanism \cite{Wolfenstein:1977ue,Mikheev:1986gs}, and ($iii$) vacuum oscillations. Even though vacuum and MSW effects are well-understood, collective oscillations are still not understood very well, even though they can drastically change the flavor-dependent neutrino fluxes. New flavor instabilities are being discovered due to spontaneously broken symmetries \cite{Raffelt:2013rqa,Abbar:2015mca,Chakraborty:2014nma,Mirizzi:2015fva,Zaizen:2020xum}, and the understanding of the so-called fast conversion \cite{Sawyer:2005jk,Sawyer:2008zs,Sawyer:2015dsa} is still very approximate \cite{Mirizzi:2015eza}. In addition, several beyond the Standard Model physics scenarios indicate that a heavy neutrino state can decay to a lighter one \cite{Berezhiani:1991vk,Fogli:1999qt,Choubey:2000an,Lindner:2001fx,Beacom:2002cb,Joshipura:2002fb,Bandyopadhyay:2002qg,Beacom:2004yd,Berryman:2014qha,Picoreti:2015ika,Frieman:1987as,Mirizzi:2007jd,GonzalezGarcia:2008ru,Maltoni:2008jr,Baerwald:2012kc,Broggini:2012df,Dorame:2013lka,Gomes:2014yua,Abrahao:2015rba,Coloma:2017zpg,Choubey:2018cfz,deSalas:2018kri,Denton:2018aml,Abdullahi:2020rge}, which again will alter the prediction of the flavor-dependent neutrino fluxes. Each of these scenarios indicate that different flavor content of neutrinos can reach terrestrial detectors, depending on which one is the correct one. 
Hence, it is of great interest to look for novel ways to test the SN neutrino flavor content. These motivate a data-driven way to identify whether the DSNB imply standard MSW or goes beyond. 

Present and upcoming neutrino detectors will have the capability to probe the flavor content of SN neutrinos. For Galactic core collapse, detectors such as Super-Kamiokande (Super-K), the Deep Underground Neutrino Experiment (DUNE), and Jiangmen Underground Neutrino Observatory (JUNO) anticipate high statistics of neutrino events. These will allow to probe, e.g., the number of neutrino species \cite{Raffelt:2011nc}, constrain the parameters of the neutrino decay \cite{Ando:2004qe} and the neutrino magnetic moment \cite{Balantekin:2007xq,deGouvea:2012hg}. Using multiple detection channels will allow flavor dependent studies. For example, the $\bar{\nu}_e$ will be reconstructed using inverse-beta decay (IBD) events, $\nu_e$ reconstructed using $\nu$-electron scattering and/or CC on argon at DUNE, and $\nu_x$ probed using $\nu$-proton scattering at JUNO \cite{Scholberg:2012id,Li:2017dbg,Li:2019qxi}. 

However, Galactic SNe are rare, happening only a few times per century \cite{Rozwadowska:2021lll}. Meanwhile, every hour a vast number of explosions happen in the whole universe. The resulting neutrinos are called the diffuse supernova neutrino background (DSNB) and have energies between a few MeV up to tens of MeV (see e.g., Refs.~\cite{Lunardini:2010ab,Beacom:2010kk} for recent reviews). The DSNB complements the Galactic supernova searches (and does not replace them) since it traces the mean neutrino emission; a Galactic search will need centuries to collect a reasonable sample size of core collapses. 
Although the DSNB has not yet been detected,  the gadolinium (Gd) enrichment at Super-K, originally proposed in Ref.~\cite{Beacom:2003nk}, will drastically reduce the backgrounds making it possible to observe a few DSNB $\bar\nu_e$ events per year \cite{Zhang:2013tua}. Furthermore, long baseline neutrino experiments which are in construction will be sensitive to different flavors of DSNB neutrinos with high statistics, e.g., Hyper Kamiokande (Hyper-K) experiment in Japan \cite{Abe:2018uyc}, the JUNO in China \cite{An:2015jdp} and DUNE in the United States \cite{Abi:2020evt}. 

In this work we study the flavor structure of the DSNB, and quantify how well future neutrino detectors will determine the ratio of the flux of $\nu_\alpha$ divided by the total flux, $f_\alpha$, for different oscillation scenarios. For detectors, we consider
DUNE, Hyper-K and JUNO, and we use these to find the upper/lower bounds on the flavor ratios $f_\alpha$ for $\alpha= \nu_e, \bar{\nu}_e, \, {\rm and} \, \nu_x$. We consider multiple detection channels: IBD, absorption on argon, $\nu$-electron scattering, and $\nu$-proton scattering. We employ different neutrino emission models and allow a generous uncertainty on the DSNB flux,
and we compare the ranges of $f_\alpha$ adopting the MSW mechanism in the absence and presence of neutrino decay.   
The paper is organized as follow:  In Sec.~\ref{sec:flux}, we describe the simulations used to describe the DSNB flux. In Sec.~\ref{sec:experiments}, we discuss the calculation of the DSNB events at several different experiments, namely DUNE, Hyper-K and JUNO. We show our results in Sec.~\ref{sec:results} and finally, we make our concluding remarks in Sec.~\ref{sec:conslusion}.

\section{DSNB flux}\label{sec:flux}

\subsection{Neutrino emission models}

For the calculation of the DSNB fluxes we employ two approaches. In the first, we adopt the predictions of Ref.~\cite{Horiuchi:2017qja}, hereafter H18, where the authors performed a detailed study combining multiple simulations of both core collapse of massive stars to neutron stars and black holes. This approach is an attempt to account for the progenitor-dependence of core-collapse neutrino emission. In the second approach, we adopt a simple Fermi-Dirac (FD) model described by a thermal temperature. 

In H18, suites of axisymmetric two-dimensional hydrodynamical simulation of core collapse to neutron stars were combined with half a dozen spherically symmetric one-dimensional simulations of core collapse to black holes. For the collapse to neutron stars, the authors considered two sets of simulations: ($i$) a large suite of over 100 progenitors \cite{Nakamura:2014caa} using a leakage approximation for heavy lepton neutrino transport and a more sophisticated ray-by-ray neutrino transport for $\nu_e$ and $\bar{\nu}_e$, and ($ii$) a smaller suite of 18 progenitors \cite{Summa:2015nyk} using a ray-by-ray neutrino transport for all neutrino flavors. Both were also augmented by a simulation of core collapse of an ONeMg core star \cite{Huedepohl:2009wh}. For the purposes of this study, we adopt the predictions based on the latter set, since we need accurate spectral information for all neutrino flavors.

Here, we briefly summarize the procedure and refer the reader to H18 for details. For each progenitor, the neutrino’s spectral parameters (i.e., the total energy, mean energy, and spectral shape or pinching parameter) are obtained directly from the simulations. Since simulations cover only the first $\sim 1$ second after core bounce, a time extrapolation is necessary to cover the long-term neutrino emission. Such an extrapolation is not needed for collapse to black holes, since the simulations cover the time until black hole formation. The time-summed neutrino spectrum is then fit to a pinched FD functional form as  \cite{Keil:2002in,Tamborra:2012ac},
\bea
F(E)=\frac{(1+\langle \alpha \rangle)^{(1+\langle \alpha \rangle)}}{\Gamma(1+\langle \alpha \rangle)}\frac{E_\nu^{\rm{tot}}E^{\langle \alpha \rangle}}{\langle E_\nu \rangle^{2+\langle \alpha \rangle}}\exp\Big[-(1+\langle \alpha \rangle)\frac{E}{\langle E_\nu \rangle}\Big],
\eea
where $E_\nu^{\rm{tot}}$ is the total neutrino energy emitted from the SNe, $\langle E_\nu \rangle$ is the  mean neutrino energy and $\langle \alpha \rangle$ is a spectral shape parameter. 
Next, the mean neutrino spectrum $\frac{dN}{dE}$ for a population of progenitors is computed by weighting each progenitor by the initial mass function (IMF),
\bea
\frac{dN}{dE}=\sum_i\frac{\int_{\Delta M_i} \psi(M) dM}{\int_8^{100} \psi(M) dM}F_i(E),
\eea
where $\psi(M)=dn/dM$ is the IMF of stars and $\Delta M_i$ is the mass range assigned to the $i$th progenitor. We use a Salpeter IMF which is $\psi(M)\propto M^\eta$, where $\eta=-2.35$.  The integration is performed using the solar-metallicity progenitors of Ref.~\cite{Woosley:2002zz} augmented by an ONeMg core star \cite{Nomoto:1984,Nomoto:1987} to represent the mass range of core-collapse progenitors taken to be between $(8-100)~M_\odot$. 

The contribution from core collapse to black holes is uncertain due to the unknown fraction of progenitors that collapse to black holes. Theoretically, this is a complex prediction impacted by the progenitor structure, the physics of hot dense nuclear matter, as well as unsettled aspects of the SN explosion mechanism (e.g., \cite{OConnor:2010moj,Ertl:2016sag,Ertl:2015rga,Sukhbold:2015wba,Ertl:2019zks}). However, recent observations suggest the fraction may be fairly large. For example, searches for the disappearance of massive stars \cite{Kochanek:2008mp} lead to a fraction of failed explosions of 4--43\% percent at 90\% C.L.~\cite{Gerke:2014ooa,Adams:2016hit,Adams:2016ffj}, which is not only consistent with other observables but can also explain puzzles related to SNe and their remnants \cite{Horiuchi:2011zz,Yuksel:2012zy,Kochanek:2013yca,Horiuchi:2014ska,Kochanek:2014mwa}.
A simple procedure to include the uncertain contribution from black holes is to parametrize the fraction of stars that collapse to black holes (e.g., \cite{Lunardini:2009ya,Moller:2018kpn}). In this work, we opt instead to adopt a fixed black hole fraction of $17$\% (using the parameterization of H18, this corresponds to a ``critical compactness'' for progenitors to collapse to black holes of $\xi_{2.5}=0.2$) and attribute a generous uncertainty to the DSNB flux when performing our statistical forecasts.

In the second approach, we describe the mean supernova neutrino spectrum effectively by using a thermal FD distribution \cite{Beacom:2010kk},
\bea
\frac{dN}{dE}=\frac{E_\nu^{\rm{tot}}}{6}\frac{120}{7\pi^4}\frac{E^2}{T_\nu^4}\frac{1}{e^{E/T\nu}+1},
\eea
where $T_\nu$ is the temperature of each neutrino flavor, $E_\nu^{\rm{tot}} = 3 \times 10^{53}$ erg is the total energy liberated, and the factor $1/6$ represents equipartition into the 6 neutrino and anti-neutrinos. The temperatures can be obtained by fitting the neutrino emissions predicted by core-collapse simulations, and generally show the following hierarchy: $T_{\nu_e}<T_{\bar\nu_e}<T_{\nu_x}$. Compilations generally cluster around $T_{\nu_e} \approx 3$--5 MeV, $T_{\bar\nu_e} \approx 4$--6 MeV, and $T_{\nu_x} \approx 4$--7 MeV \cite{Horiuchi:2008jz,Mathews:2014qba}. While this approach has often been used in the past literature, it does not take into account the progenitor dependence of the neutrino emission which is now known to be substantial \cite{Horiuchi:2017qja,Kresse:2020nto}. The benefit however is its simplicity and ease of parameterization. In the following, we therefore consider both approaches.

\subsection{DSNB formulation}\label{sec:dsnb}

The DSNB differential flux is the integrated neutrino flux over redshift, appropriately weighted by the core-collapse rate, given by
\bea\label{eq:DSNBflux}
\frac{d\phi^0}{dE}=c\int_0^{z_{max}} R_{CC}(z)\frac{dN}{dE^\prime}(1+z)\Big|\frac{dt}{dz}\Big|dz,
\eea
where $c$ is the speed of light, $z$ is the redshift, $\frac{dN}{dE^\prime}$ is the mean neutrino spectrum per core collapse, $E^\prime=E(1+z)$, and  $|{dt}/{dz}|=H_0(1+z)[\Omega_m(1+z)^3+\Omega_\Lambda]^{1/2}$, where $H_0=70$~${\rm{kms}}^{-1}{\rm{Mpc}}^{-1}$ is the Hubble parameter, while $\Omega_m =0.3$ and $\Omega_\Lambda = 0.7$ are the matter and vacuum energy densities, respectively. The upper index in $\phi^0$ refers to fluxes which are calculated at the source (pre-oscillation). We take $z_{max}\equiv 5$, which is large enough to incorporate the majority of the DSNB flux. We model the cosmic history of the comoving core-collapse rate, $R_{CC}(z)$, by
\bea \label{eq:rate}
R_{CC}(z)=\dot\rho^*(z)\frac{\int_8^{100}\psi(M) dM}{\int_{0.1}^{100}M\psi(M) dM}
\eea
where $\psi(M)$ is once again the IMF and $\rho^*(z)$ is the rate of the cosmic star formation. 
The cosmic star-formation rate is derived from the observed luminosity density of various star-forming priors, which are converted through use of appropriate conversion factors \cite{Kennicutt:1998zb,Hopkins:2004ma}. There are various compilations of the cosmic star-formation rate and functional fits through subsets of the available data (see, e.g., \cite{Hopkins:2006bw,Madau:2014bja}). We adopt the smoothed piece-wise form of Ref.~\cite{Yuksel:2008cu} (see Refs.~\cite{Hopkins:2006bw,Yuksel:2008cu,Horiuchi:2008jz} for numerical values). While the cosmic star-formation rate has significant systematic uncertainty, one of the most dominant---the IMF---fortunately mostly cancels by performing the product with the integration ratio in Eq.~(\ref{eq:rate}). The remaining uncertainty is dominated by the scatter between the measurements and is generally in the range of $\sim 20$\% over redshifts of importance for the DSNB (see discussions in, e.g., Refs.~\cite{Hopkins:2006bw,Horiuchi:2008jz,Madau:2014bja,Mathews:2014qba}). 

\begin{figure}[h!]
\centering
\includegraphics[width=0.6\textwidth]{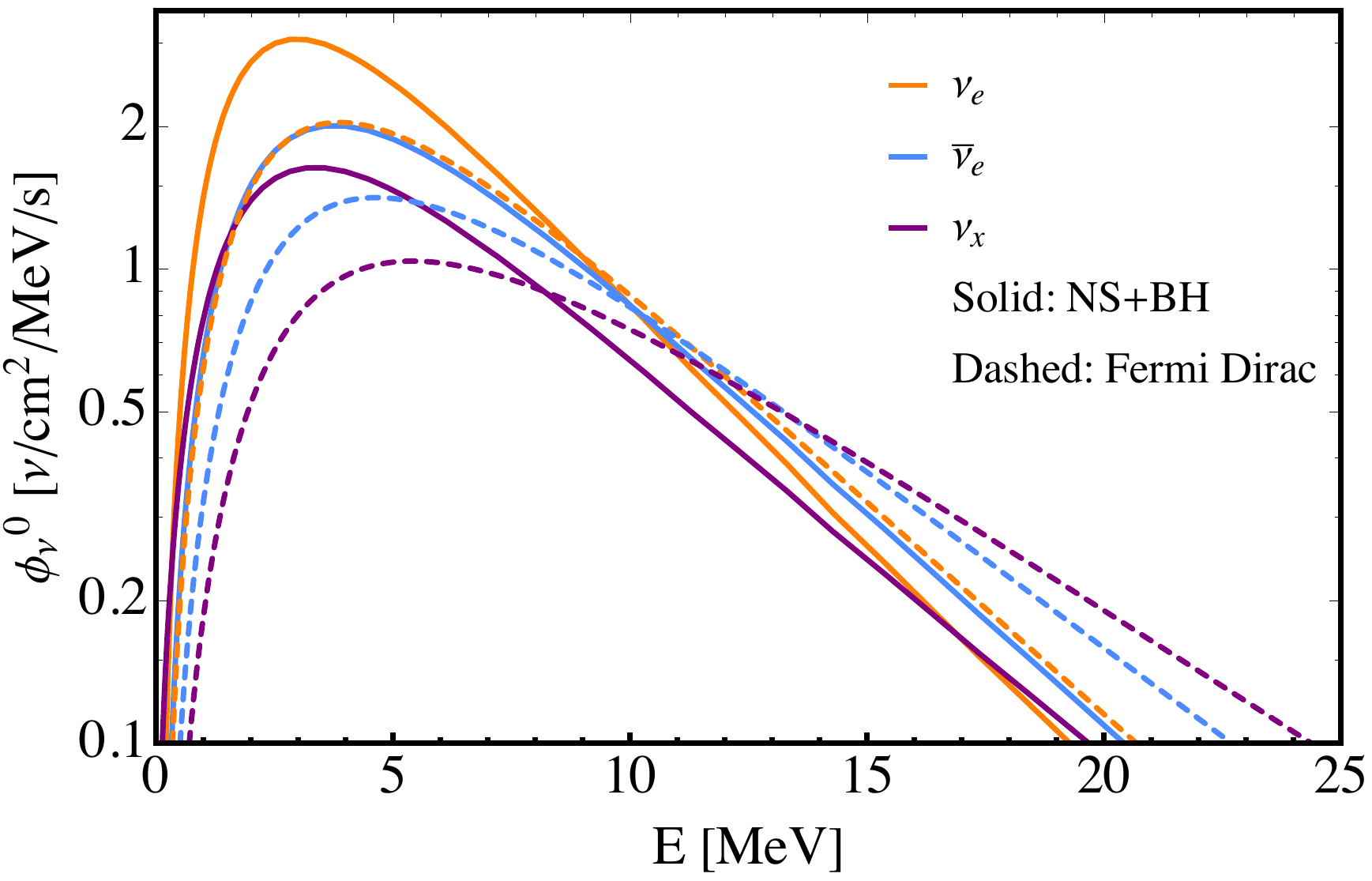}
\caption{\label{fig:flux0} The DSNB flux without any oscillation effects for $\nu_e$ (orange), $\bar\nu_e$ (blue) and $\nu_x$ (purple) respectively. The solid curves represent the 83\% neutron star plus 17\% black hole (NS+BH) of H18, while the temperature dependant FD distributions are shown with dashed curves. 
For the thermal FD distribution we have chosen $T_{\nu_e}=5$~MeV, $T_{\bar\nu_e}=6$~MeV and $T_{\nu_x}=7$~MeV, respectively. 
}
\end{figure}

Figure \ref{fig:flux0} shows the differential DSNB flux as a function of the neutrino energy for $\nu_e$, $\bar\nu_e$ and $\nu_x$ in orange, blue and purple, respectively. The predictions of H18 containing a black hole fraction of 17\% are shown with solid curves. For comparison we also show the fluxes we have calculated using the thermal FD flux with the dashed curve, where we have chosen $T_{\nu_e}=5$~MeV, $T_{\bar\nu_e}=6$~MeV and $T_{\nu_x}=7$~MeV, respectively. These thermal FD fluxes are smaller than those of H18 at a few MeV but dominate above $\sim 12$ MeV, giving significantly higher neutrino rates in that region.  

\subsection{MSW}
Once neutrinos are emitted from the neutrinospheres they will oscillate during their propagation before reaching detectors on Earth. 
The fluxes after oscillation are therefore a combination of the initial fluxes emitted from the neutrinospheres. 
During propagation through the progenitor envelope, the Mikheev-Smirnov-Wolfenstein (MSW) effect becomes important due to the coherent scattering of neutrinos on electrons.  
Due to the matter potential, electron neutrinos $\nu_e$ will exit as $\nu_3$ while the other states $\nu_x$ will exit as $\nu_1$ and $\nu_2$, in the normal mass hierarchy (NH). In the anti-neutrino sector, electron anti-neutrinos $\bar\nu_e$ exit as $\bar\nu_1$, while $\bar\nu_x$ exit as $\bar\nu_2$ and $\bar\nu_3$. Therefore, we may assume the following relations between the temperatures of the flavor and mass eigenstates: 
\bea
T_{\nu_3}=T_{\nu_e},\quad T_{\nu_1}=T_{\nu_2}=T_{\nu_x},\quad T_{\bar\nu_1}=T_{\bar\nu_e},\quad T_{\bar\nu_2}=T_{\bar\nu_3}=T_{\nu_x},
\eea
again for NH. If the matter density along the neutrino trajectory varies slowly (the adiabatic propagation) then neutrinos and anti-neutrinos arrive at terrestrial detectors with the same mass eigenstate as they were in SNe, and we can write $\phi_\alpha = \sum_i |U_{\alpha i}|^2 \phi^0_i$, where  $U$ represents the Pontecorvo-Maki-Nakagawa-Sakata (PMNS) matrix. This yields the fluxes at terrestrial detectors \cite{Dighe:1999bi,Lu:2016ipr}
\bea 
\phi_{\nu_e}&=&\phi^0_{\nu_x},\\
\phi_{\bar\nu_e}&=&c_{12}^2\phi^0_{\bar\nu_e}+s_{12}^2\phi^0_{\nu_x},\\
\phi_{\nu_x}&=&\frac{1}{4}\Big[\phi^0_{\nu_e}+s_{12}^2\phi^0_{\bar\nu_e}+(2+c_{12}^2)\phi^0_{\nu_x}\Big]
\eea 
for the normal hierarchy (NH) and 
\bea 
\phi_{\nu_e}&=&s_{12}^2\phi^0_{\nu_e}+c_{12}^2\phi^0_{\nu_x},\\
\phi_{\bar\nu_e}&=&\phi^0_{\nu_x},\\
\phi_{\nu_x}&=&\frac{1}{4}\Big[c_{12}^2\phi^0_{\nu_e}+\phi^0_{\bar\nu_e}+(2+s_{12}^2)\phi^0_{\nu_x}\Big]
\eea 
for the inverted hierarchy (IH), where $\phi^0_{\nu_\alpha}$ is the initial flux of (anti)neutrino flavor $\nu_\alpha$, $c_{12}^2=1-s_{12}^2$ and $s_{12}^2=0.310$ is the solar mixing angle. In this paper we assume the NH for the mass ordering and show the results for this case. The results for the IH are very similar. 

Collective flavour oscillations also occur within a few hundred kilometers from the core due to $\nu-\nu$ coherent scattering \cite{Duan:2006an,Hannestad:2006nj,Fogli:2007bk,Dasgupta:2009mg,EstebanPretel:2007ec,Dasgupta:2008my,Dasgupta:2011jf,Chakraborty:2015tfa,Dasgupta:2016dbv,Izaguirre:2016gsx,Capozzi:2018clo}. While these collective oscillations are currently the focus of much research, it is likely to be time dependent, meaning they will undergo some level of averaging out for the DSNB which detects the time-integrated neutrino spectrum. Also, during the late cooling phase of the protoneutron star, when some half of the neutrino emission occurs, the difference in spectra between flavors is much reduced, implying oscillation effects will not be pronounced. It has been predicted that collective neutrino oscillations would be subdominant compared to MSW and have an effect of less than $\sim 10\%$ \cite{Chakraborty:2008zp}. However, the results after collective neutrino mixing, including its time and energy dependence, remain highly uncertain. Therefore, we do not include the uncertain collective neutrino oscillation effects in our analysis. Instead, we will estimate how well deviations from MSW can be probed with future datasets.

\subsection{Neutrino Decay}\label{sec:decay}
In principle massive neutrinos can decay to the lighter states. Within the SM this requires the lifetime to be larger than the age of the universe. However, neutrinos can decay faster if there are interactions beyond the standard model.  We follow the decay mechanism described in Ref.~\cite{deGouvea:2020eqq}. 
Taking the NH to be the correct one, the third mass eigenstate can decay to $\nu_1$.  Hence, this can significantly change the flavor content of the DSNB flux arriving to the earth. 
We assume that neutrinos are Majorana states and they can interact with a massless scalar $\phi$ within the following Lagrangian:
\bea
\mathcal{L} \supset \frac{g}{2} {\nu}_i\nu_j \phi+{\rm{h.c.}}.
\eea
We are interested in the case where $\nu_3$ with mass $m_3$ is produced as a left handed particle, where it can decay to $\nu_1$ as a left handed (helicity conserving) or right handed (helicity flipping) daughter particle. In the lab frame the decay width of these two processes are the same:
\bea 
\Gamma(E_3)=\frac{g^2 m_3^2}{32\pi E_3}=\frac{1}{E_3}\frac{m_3}{\tau_3},
\eea
where $\tau_3$ is the lifetime. We can then calculate the flux of the mass eigenstates at the Earth as:
\bea 
{\phi_{\nu_3}}(E)&=&c\int_0^{z_{max}} R_{CC}(z)\frac{dN_{\nu_e}}{dE^\prime}(1+z)\Big|\frac{dt}{dz}\Big| e^{-\Gamma(E)\zeta(z)}dz,\\
{\phi_{\bar\nu_3}}(E)&=&c\int_0^{z_{max}} R_{CC}(z)\frac{dN_{\nu_x}}{dE^\prime}(1+z)\Big|\frac{dt}{dz}\Big| e^{-\Gamma(E)\zeta(z)}dz,\\
{\phi_{\nu_1}}(E)&=&{\phi^0_{\nu_x}}(E)+\int_0^{z_{max}}dz\frac{1}{H(z)}\int_{{E'}}^\infty dE_3 \Gamma(E_3) \Big[\frac{d\phi_{\bar\nu_3}}{d E_3}\psi_{h.c.}(E_3,{E'})+\frac{d\phi_{\nu_3}}{dE_3}\psi_{h.f.}(E_3,{E'})
\Big] ,\nonumber\\
{\phi_{\bar\nu_1}}(E)&=&{\phi^0_{\bar\nu_e}}(E)+\int_0^{z_{max}}dz\frac{1}{H(z)}\int_{{E'}}^\infty dE_3\Gamma(E_3) \Big[\frac{d\phi_{\nu_3}}{d E_3}\psi_{h.c.}(E_3,{E'})+\frac{d\phi_{\bar\nu_3}}{dE_3}\psi_{h.f.}(E_3,{E'})
\Big] ,\nonumber
\eea
where $\zeta(z)=\int_0^z dz^\prime H^{-1}(z^\prime)(1+z^\prime)^{-2}$, we have $E'=E(1+z)$ and ${\phi^0_{\nu_\alpha}}$ expressions are given in Eq.~(\ref{eq:DSNBflux}). The other mass eigenstates are unaffected by the decay, e.g., ${\phi_{\nu_2}}(E)={\phi^0_{\nu_x}}(E)$. Finally, the energy distribution of the daughter particle for the helicity conserving and helicity flipping cases are given by
\bea 
\psi_{h.c.}(E_3,E_1)=\frac{2E_1}{E_3^2},\quad \psi_{h.f.}(E_3,E_1)=\frac{2}{E_3}\left(1-\frac{E_1}{E_3}\right). 
\eea
By inclusion of the neutrino decay the flux of $\nu_\alpha$ at the earth changes dramatically. In this case the spectrum of different flavors at the earth for the NH are given by:
\bea 
\phi_{\nu_e}^{\rm{decay}}&=&c_{12}^2\phi_{\nu_1}+s_{12}^2\phi_{\nu_2},\nonumber\\
\phi_{\bar\nu_e^{\rm{decay}}}&=&c_{12}^2\phi_{\bar\nu_1}+s_{12}^2\phi_{\bar\nu_2},\\
\phi_{\nu_x}^{\rm{decay}}&=&\frac{1}{4}\Big[s_{12}^2(\phi_{\nu_1}+\phi_{\bar\nu_1})+c_{12}^2(\phi_{\nu_2}+\phi_{\bar\nu_2})+\phi_{\nu_3}+\phi_{\bar\nu_3}\Big]\nonumber,
\eea
where similar expressions can be obtained for the IH case. 

One can be sensitive to neutrino decay of $\tau/m\sim10^5$~s/eV from the neutronization burst of galactic SN \cite{deGouvea:2019goq}. The DSNB however, due to the further distances, can be sensitive to the neutrino lifetimes of $\tau/m\sim10^{10}$~s/eV. It was shown in Ref.~\cite{deGouvea:2020eqq} that using the data of Hyper-K doped with Gd as well as THEIA one can reach a $3\sigma$ sensitivity of $\tau_3/m_3\lesssim 5\times10^9$ after 20 years of data taking. Ref.~\cite{Fogli:2004gy} has also studied DSNB neutrino decay, but they have found one order of magnitude smaller sensitivity.

\section{DSNB rate at different detectors}\label{sec:experiments}
Since the DSNB flux is extremely small and only concentrates around a few MeV, enormous neutrino detectors with the ability to distinguish between the DSNB and different sources of relevant background are needed. The next generation neutrino experiments are expected to be able to deliver this. In this work we consider three different experiments which each will detect different DSNB flavors: DUNE will detect $\nu_e$ through the CC interaction on liquid Argon (LAr) while Hyper-K and JUNO will be sensitive to $\bar\nu_e$ through the IBD scattering. The sensitivity to heavy lepton neutrinos is more limited, but the elastic neutrino electron scattering channel at Hyper-K will be a source for detecting the sum of all flavors. Furthermore, sensitivity exists through proton scattering at JUNO, provided backgrounds can be sufficiently mitigated. The next subsections are devoted to the experimental details of each of these detectors.  We list the experimental details in Table~\ref{tab:experiments}. 

\begin{table}[h]
\begin{center}
\scalebox{0.9}{
\begin{tabular}{|c|c|c|c|c|}
\hline\hline
\bf Experiment& \bf Fiducial Mass (kt) & \bf Targets & \bf Energy range (MeV) & \bf efficiency
\\\hline\hline
DUNE & 40 & $6.02\times10^{32}$ & $19-32$ & $86\%$\\\hline
Hyper-K (IBD) & 374 & $2.50\times 10^{34}$ & $12-24$ & 67\%\\\hline
Hyper-K ($\nu-e$) & 374 & $1.25\times10^{35}$ & $10-20$ & 100\%\\\hline
JUNO (IBD) & 17 & $1.21\times 10^{33}$ & $10-22$ & $50\%$\\\hline
JUNO ($\nu-p$) & 17 & $1.21\times 10^{33}$ & $0.2-1.5$ & $100\%$\\\hline\hline
\end{tabular}}
\end{center}
\caption{\label{tab:experiments} Summary of the detectors set-up and values assumed in our calculations.}
\end{table}

\begin{figure}[h!]
\centering
\includegraphics[width=0.6\textwidth]{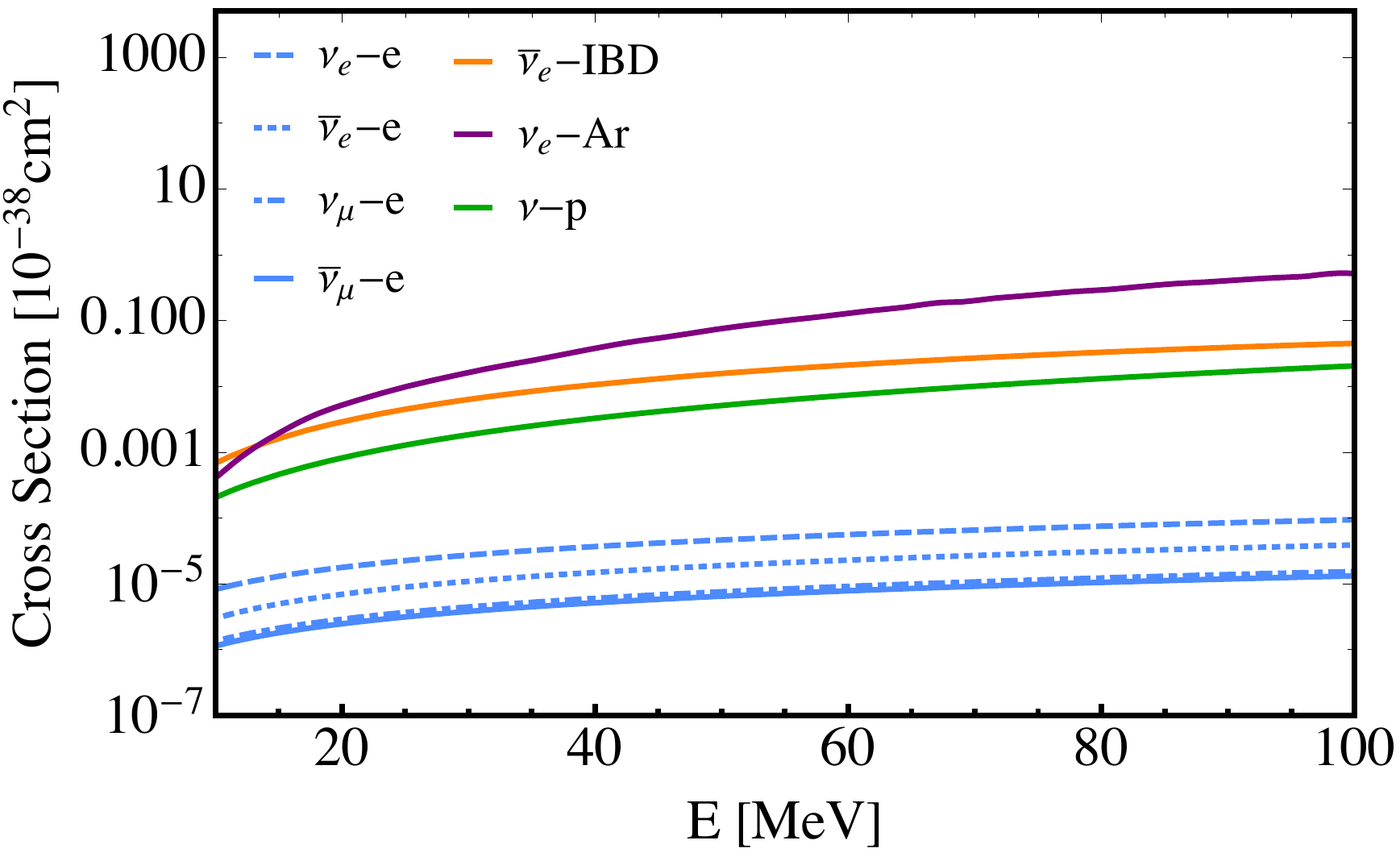}
\caption{\label{fig:xsection} The cross sections as a function of the neutrino energy for the IBD scattering of $\bar\nu_e$ (orange), the CC $\nu_e$ scattering on Argon (purple) and the neutrino-proton elastic scattering (green). The elastic scattering of different neutrino flavors on electron are also shown in blue. }
\end{figure}

\subsection{DUNE}

Using the LAr detector of DUNE, we will be able to detect the electron component of the DSNB flux via the CC interaction $\nu_e+Ar\rightarrow e^-+K^+$ \cite{Abi:2020evt}. The signature of this process is observing an electron accompanied by the decay products of the excited $K^*$. 
The DUNE far detector which has a fiducial mass of at least $40$~kt will be sensitive to DSNB neutrinos from around 5 MeV to a few tens of MeV. The main sources of background at this energy are the solar hep neutrinos which have an endpoint of $18.8$ MeV, and the atmospheric flux of electron neutrinos which rises in an energy around $40$ MeV. Since the exact details of the DSNB backgrounds at DUNE are still under investigation, we assume DUNE will have a background similar to ICARUS \cite{Moller:2018kpn,Cocco:2004ac}. We calculate the events in the range of $19-31$ MeV to get rid of these backgrounds. We take the CC cross section of neutrinos on Argon from Ref. \cite{Abi:2020evt} which we have shown in Fig.~\ref{fig:xsection}.

\begin{figure}[h!]
\centering
\includegraphics[width=1\textwidth]{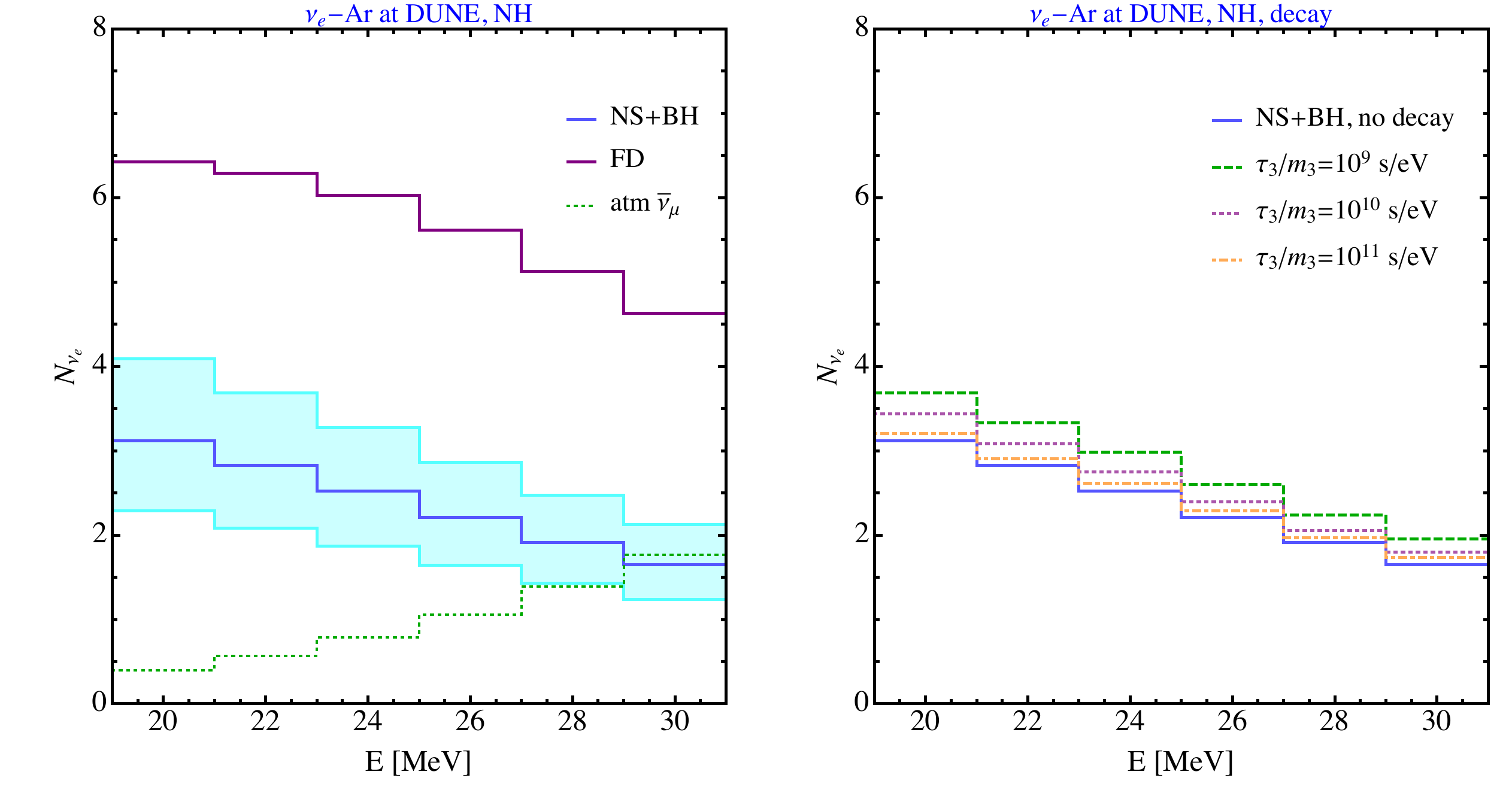}
\caption{\label{fig:DUNE}  Expected DSNB event rates as a function of the neutrino energy at the DUNE. Left panel: the rates for the NS$+$BH model of H18 and the thermal FD model are shown in dark blue and purple, respectively. For the thermal FD model we have chosen temperatures of $T_{\nu_e}=5$~MeV, $T_{\bar\nu_e}=6$~MeV and $T_{\nu_x}=7$~MeV, respectively. The light blue band shows the variability of the H18 flux within the theoretical uncertainties described in Section \ref{sec:dsnb}. The background is shown in dotted green curve. Right panel: comparison of the event rates for the H18 model considering MSW scenario without decay and with decay for different values of $\tau_3/m_3$. }
\end{figure}

The number of expected DSNB events at the DUNE far detector is calculated by
\bea \label{eq:DUNEevents}
N_i^{\rm{DUNE}}={n}_{{\rm{DUNE}}}\times T \times \epsilon \int_i \frac{d\phi_{\nu_e}}{dE}\sigma(E) dE
\eea
where ${n}_{{\rm{DUNE}}}$ is the total number of Argon targets at the detector mass 
and $T$ is the total lifetime of the experiment. The fiducial mass of the far detector is $40$ kt and we consider $20$ years of data taking. Therefore, we have
\bea
{n}_{{\rm{DUNE}}}=6.02\times10^{32}.
\eea
Using this factor and assuming a detector efficiency of $\epsilon=86\%$ \cite{Abi:2020evt} we find a total of $15$  and $38$ $\nu_e$ events with the H18 and thermal FD fluxes, respectively. Figure \ref{fig:DUNE} shows the expected number of events as a function of the neutrino energy
for the two different neutrino emission models. For the thermal FD flux we have chosen $T_{\nu_e}=5$~MeV, $T_{\bar\nu_e}=6$~MeV and $T_{\nu_x}=7$~MeV, respectively. As can be seen, these temperatures result in much higher expected events than H18, even beyond the variability of the flux within its possible uncertainty. This is consistent with our choice of temperature which are on the large end of those predicted by simulations.
For comparison, in the right panel we show the event rates for the decay scenario assuming different values for $\tau_3/m_3$, all for the H18 model. 
The shorter the lifetime, the faster the decay, and hence the higher the $\nu_1$ flux and higher the $\nu_e$ events on Earth; indeed, this is what we see in Fig.~\ref{fig:DUNE}.

\begin{figure}[h!]
\centering
\includegraphics[width=1\textwidth]{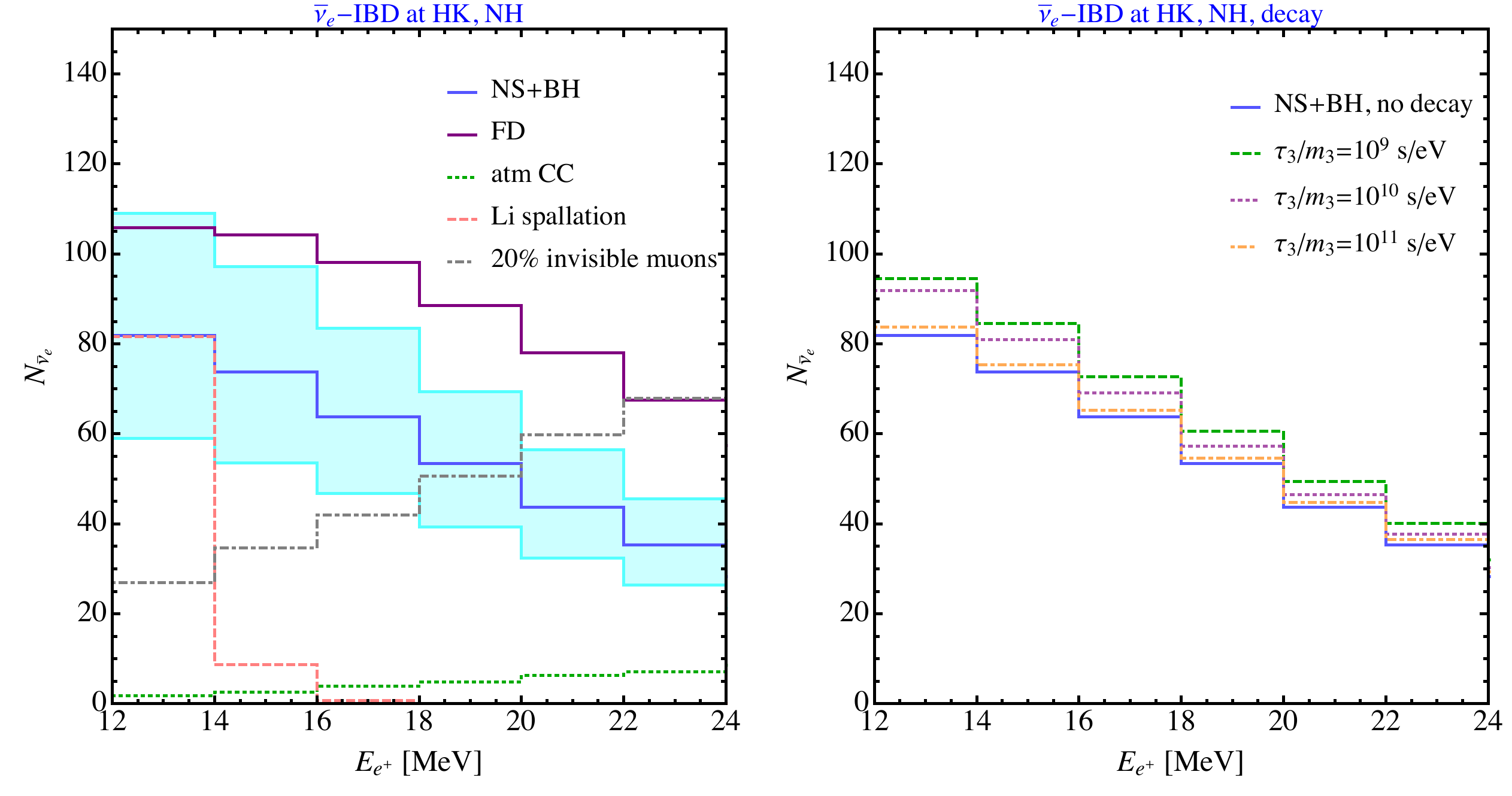}
\caption{\label{fig:HK}  Expected event rates as a function of the positron energy at Hyper-K experiment for the IBD events. Left panel: the rates for the NS$+$BH model of H18 and the thermal FD model are shown in dark blue and purple, respectively. For the temperatures of the thermal FD flux we have chosen $T_{\nu_e}=5$~MeV, $T_{\bar\nu_e}=6$~MeV and $T_{\nu_x}=7$~MeV, respectively. The light blue band shows the variability of the flux within the theoretical uncertainties. The backgrounds are shown in dotted green, dashed pink and dot dashed gray curves. Right panel: Comparing the event rates for the MSW scenario (no decay) with the decay mechanism for different values of $\tau_3/m_3$.  }
\end{figure}

\subsection{Hyper-K}

The Hyper-K experiment in Japan, which is the successor of the Super-K experiment, will be able to detect SN neutrinos down to $\sim 3$ MeV \cite{Abe:2018uyc}, although the threshold for the DSNB will be higher owing to backgrounds. It consists of two modules, each with a fiducial mass of $187$ kt. With a combined volume of 374 kt, Hyper-K will be sixteen times larger than Super-K, resulting in great sensitivity to the DSNB. Unlike the DUNE which will be sensitive to electron neutrinos, Hyper-K will detect $\bar\nu_e$ neutrinos through IBD scattering $\bar\nu_e+p\rightarrow e^++n$.  We get the IBD cross section from Ref.~\cite{Vogel:1999zy,Strumia:2003zx} (see Fig.~\ref{fig:xsection}). 

If Hyper-K is enriched with gadolinium (Gd) its sensitivity to DSNB events will be significantly improved due to the significantly reduced backgrounds. 
The CC atmospheric neutrinos and the lithium-9 spallation are subdominant backgrounds at energies around tens of MeV. On the other hand, the neutral current (NC) atmospheric neutrinos which are induced due to the $\gamma$-rays produced by the NC quasi elastic scattering, are non-negligible in this range. If we assume full tagging can be reached, we can ignore this background \cite{Moller:2018kpn}.
In practice, if the tagging efficiency will be comparable to Super-K, it will be in the 90\% range. The invisible muons will be similarly suppressed by neutron tagging but given its much higher rate it will remain as the main source of background, which increases with energy. Therefore, we consider an upper bound of $E<24$~MeV to suppress the invisible muons.  
The modeling of backgrounds at Hyper-K enriched with Gd is taken from \cite{Moller:2018kpn,Abe:2018uyc}. There are other sources of background which will not be reduced by Gd, but a cut on the energy will avoid them. The dominant background at $E<10$ MeV is the $\bar\nu_e$ from reactor neutrinos, which is however suppressed at $E>10$ MeV. Therefore, we consider the range $10$~MeV $<E<24$~MeV. 

The number of expected DSNB events through the IBD cross section at the Hyper-K detector at each bin of the positron energy is calculated by
\bea \label{eq:rateIBD}
N_i^{\rm{HK-IBD}}(E^+)={{n}}^{\rm{IBD}}_{{\rm{HK}}}\times T\times  \epsilon \int_i \frac{d\phi_{\bar\nu_e}}{dE}\sigma(E^+) dE^+,
\eea
where we have used $E=E^++\Delta$ with $\Delta=1.3$ MeV in the flux to convert the neutrino energy to the positron energy. The total number of IBD targets are $n^{\rm{IBD}}_{\rm{HK}}=2.50\times10^{34}$. Considering a tagging efficiency of 90\% and event selection efficiency of 74\%, which is slightly lower than $\sim 90$\% efficiency of the latest Super-K DSNB analysis \cite{Bays:2011si}, we adopt an overall detector efficiency of $67\%$ \cite{Abe:2018uyc}.
The Hyper-K experiment will detect a total of $444$ (747) $\bar\nu_e$ events using the H18 (thermal FD) flux. Figure \ref{fig:HK} shows the expected number of DSNB events as a function of the positron energy compared to relevant backgrounds for Hyper-K enriched with Gd. In the right panel, we also show how different values of $\tau_3/m_3$ can change the expected DSNB events under the decay scenario. Similar to $\nu_e$ events at DUNE, a faster neutrino decay can result in higher $\bar\nu_e$ events at Hyper-K.

Hyper-K can also be sensitive to the sum of all neutrino flavors  due to the elastic scattering of neutrinos on electrons: $\nu+e\rightarrow \nu+e$, where the observable will be a forward going electron. Although the cross section will be almost three orders of magnitudes suppressed compared to IBD (see Fig.~\ref{fig:xsection}), in this way Hyper-K can be sensitive to $\nu_e$ and $\nu_x$ DSNB as well. The differential $\nu-e$ scattering cross section is given by \cite{deGouvea:2019wav},
\bea
    	\frac{d\sigma}{dE_R}  = \frac{2G_{F}^2m_e}{\pi}\left\{g_1^2+g_2^2\left(1-\frac{E_R}{E}\right)^2-g_1 g_2\frac{m_e E_R}{E^2}\right\} ,
\eea
where $G_F$ is the Fermi constant, $E$ is the energy of the incoming neutrino, and $m_e$ is the electron mass and $E_R$ is its recoil kinetic energy. The couplings $g_1$ and $g_2$ for different neutrino flavors depend on the weak angle $\sin^2\theta_w$ and are listed in Table~1 of \cite{deGouvea:2019wav}. 
\begin{figure}[h!]
\centering
\includegraphics[width=1\textwidth]{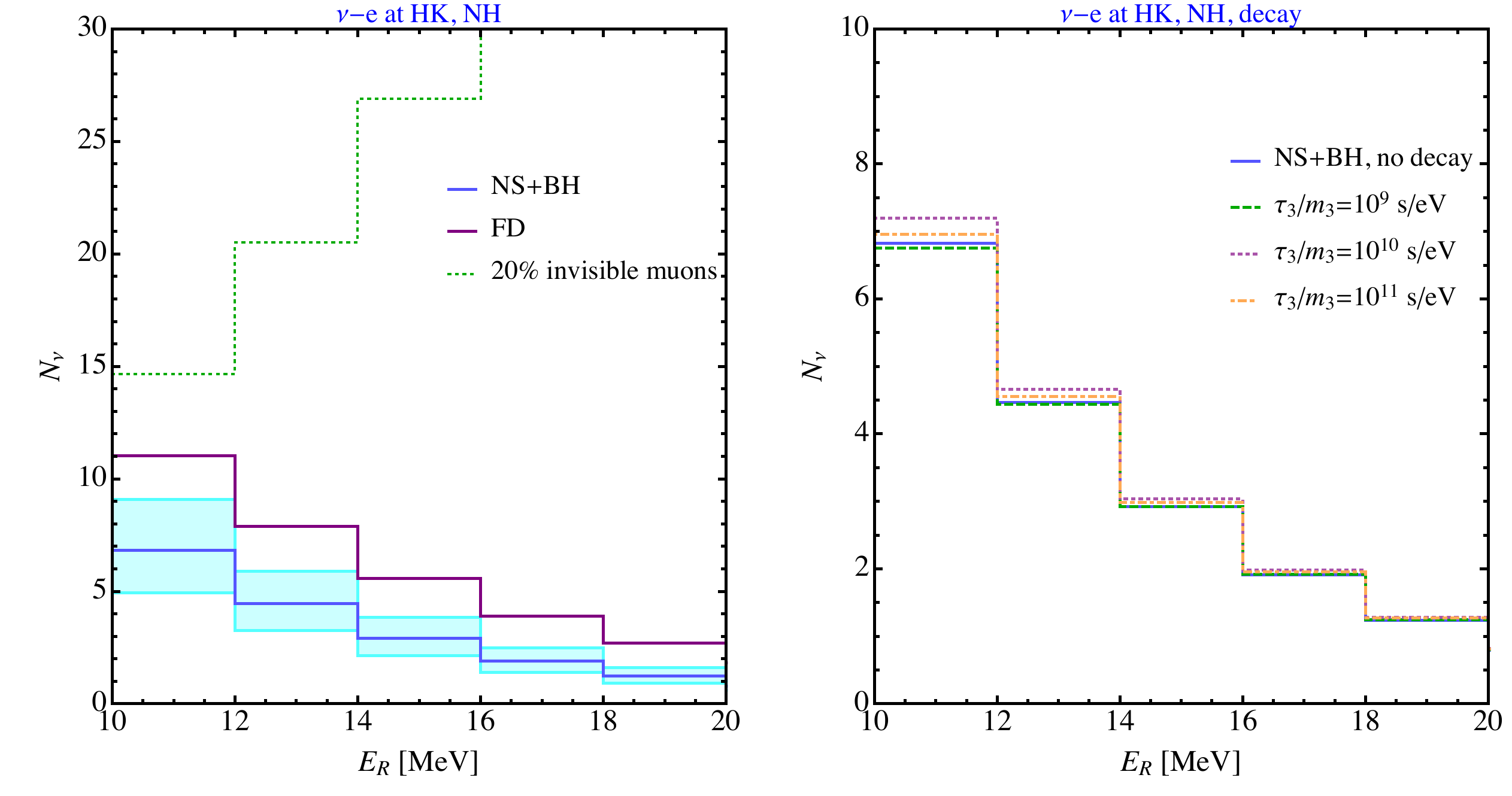}
\caption{\label{fig:HKnue}   Expected event rates as a function of the electron recoil energy at Hyper-K experiment for the $\nu-e$ events. Color coding is the same as Fig.~\ref{fig:HK}. The invisible muon background shown here is the same reduced rate as shown in Fig.~\ref{fig:HK} for clarify to show the shape of the signal, but in the analysis we use the unreduced rate.}
\end{figure}

The number of $\nu-e$ DSNB events at Hyper-K as a function of the electron recoil energy is given by
\bea \label{eq:HKnueevents}
N_i^{\rm{HK-\nu- e}}(E_R)=\sum_{\alpha=\nu_e,\bar\nu_e,\nu_x}{{n}}^{\rm{\nu-e}}_{{\rm{HK}}}\times T\times  \epsilon \int_i dE_R \int_{E_R}^{E_R^{\rm{max}}}dE \frac{d\phi_{\nu_\alpha}}{dE}\frac{d\sigma_{\nu_\alpha}}{dE_R},
\eea
where the number of electron targets is ${{n}}^{\nu-e}_{{\rm{HK}}}=1.25\times10^{35}$. Considering the recoil energy larger than 10 MeV we can get rid of the reactor neutrinos, and the solar angular cut can get rid of the solar neutrinos as the main sources of background. The invisible muons are however unavoidable, for which we show in Fig.~\ref{fig:HKnue} the results from Ref.~\cite{Beacom:2003nk}. Note however that this is the reduced rate considering coincidence tagging which applies to IBD events. For $\nu-e$ events the coincidence cut does not apply, and we have to remove the background reduction factor of five.  Assuming a detection efficiency of $100\%$ above the threshold, we find that Hyper-K will be able to observe $17~(31)$ $\nu-e$ events using the H18 (thermal FD) flux, in the range of $10$~MeV$<E_R<20$~MeV, but this is small compared to the invisible muon events. Thus, Hyper-K will not be able to claim standalone detection, but would constrain extremely high fluxes of $\nu_e$ and $\nu_x$ when combined with its IBD channel (and when further combined with DUNE's measurement of $\nu_e$, constrain high $\nu_x$ fluxes). We  show in Fig.~\ref{fig:HKnue}
the expected number of $\nu-e$ DSNB events at Hyper-K as a function of the electron recoil energy (left panel), as well as the comparison between the MSW with and without decays (right panel).

\subsection{JUNO}
\begin{figure}[h!]
\centering
\includegraphics[width=1\textwidth]{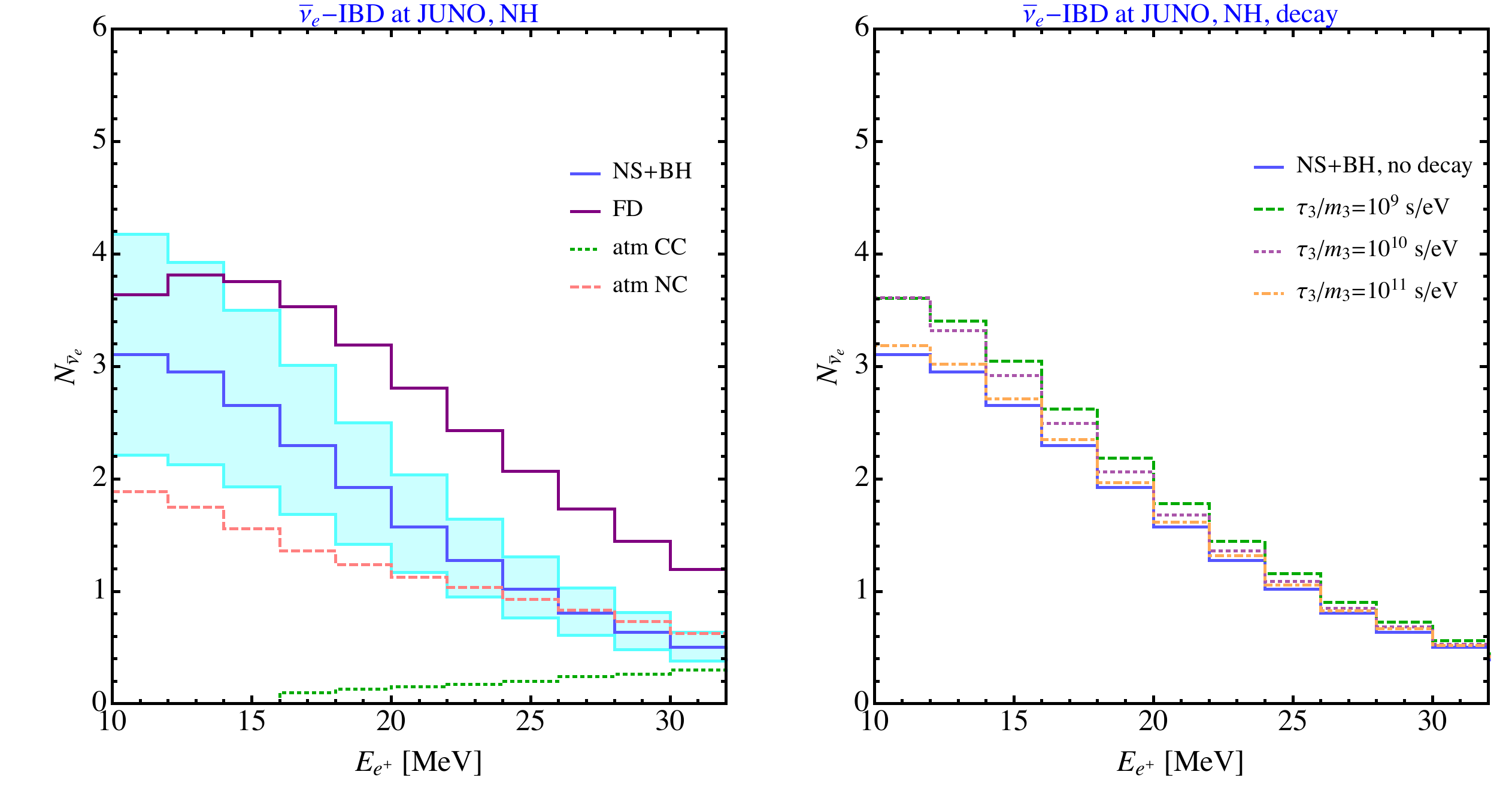}
\caption{\label{fig:JUNOIBD} Number of IBD events at JUNO as a function of the positron energy. The color coding is similar to Fig.~\ref{fig:HK}.  }
\end{figure}

The JUNO detector in China has a fiducial volume of 17 kt, made of linear alkylbenzene liquid scintillator (C$_6$H$_5$C$_{12}$H$_{25}$) \cite{An:2015jdp}. JUNO will be able to detect $\bar\nu_e$ events thought IBD interaction. We calculate the number of IBD events using Eq.~(\ref{eq:rateIBD}) with ${n}^{\rm{IBD}}_{{\rm{JUNO}}}=1.21\times10^{33}$ and a detection efficiency of $50\%$. The main sources of background here are the reactor electron anti-neutrinos, which we can get rid of with a cut of $E_{e^+}>10$ MeV, and the atmospheric $\bar\nu_e$ events as well as the NC atmospheric, which we take from Refs.~\cite{An:2015jdp,Moller:2018kpn}. We show in Fig.~\ref{fig:JUNOIBD} the expected number of IBD events at JUNO as a funciton of the positron energy as well as the backgrounds. In total we expect JUNO will be able to detect $19~(30)$ IBD events in 20 years of data taking, using the H18 (thermal FD) flux, in the range of $10$~MeV$<E_{e^+}<32$~MeV.

\begin{figure}[h!]
\centering
\includegraphics[width=1\textwidth]{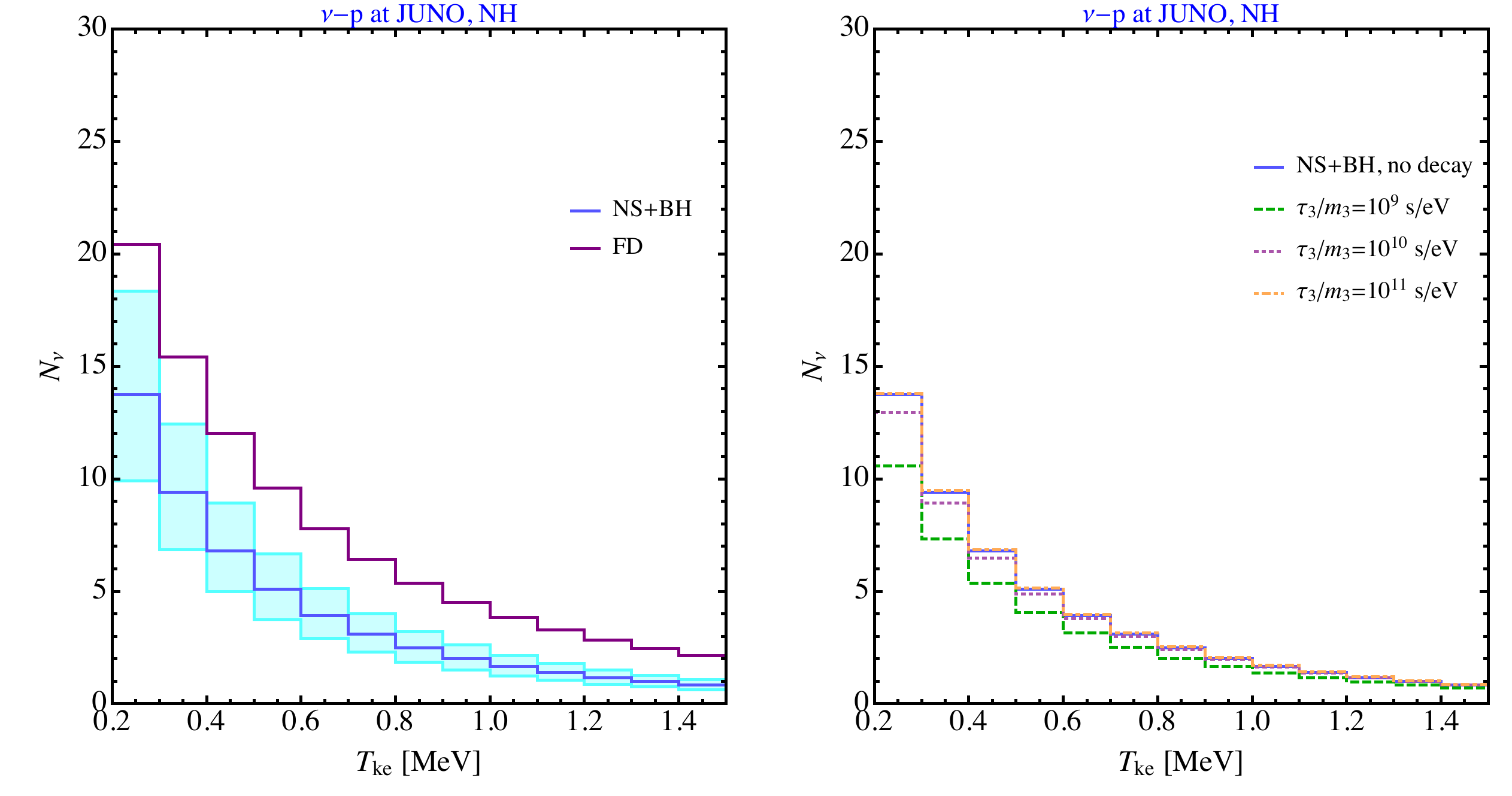}
\caption{\label{fig:JUNOnup} Number of $\nu-p$ events at JUNO as a function of the kinetic energy of proton. The color coding is the same as Fig.~\ref{fig:HK}. }
\end{figure}

In addition to the IBD detection of $\bar\nu_e$, JUNO can also detect neutrino events with the neutrino-proton elastic scattering which is sensitive to the sum of all neutrino flavors. In this case the differential cross section is given by \cite{Dasgupta:2011wg}
\bea
    	\frac{d\sigma}{dT_{\rm ke}} & =& \frac{G_{F}^2m_p}{\pi}\left\{c_v^2\left(1-\frac{m_p T_{\rm ke}}{2E^2}\right)^2+c_a^2\left(1+\frac{m_p T_{\rm ke}}{2E^2}\right)^2\right\} \nonumber\\
    	&=&4.83\times10^{-42}\frac{{\rm{cm}}^2}{{\rm{MeV}}}\left(1+466\frac{T_{\rm ke}}{E^2}\right),
\eea
where $m_p$ is the proton mass, $T_{\rm ke}$ is its kinetic energy, $c_v=0.04$ and $c_a = 1.27/2$. The number of $\nu-p$ events at JUNO is given by
\bea \label{eq:JUNOnupevents}
N_i^{\rm{JUNO-\nu- p}}(T)=\sum_{\alpha=\nu_e,\bar\nu_e,\nu_x}{{n}}^{\rm{\nu-p}}_{{\rm{JUNO}}}\times T\times  \epsilon \int_i dT_{\rm ke} \int_{\sqrt{m_p T_{\rm ke}/2}}^{T_{\rm ke}^{\rm{max}}}dE \frac{d\phi_{\nu_\alpha}}{dE}\frac{d\sigma}{dT_{\rm ke}},
\eea
where ${{n}}^{\rm{\nu-p}}_{{\rm{JUNO}}}=1.21\times10^{33}$ (the same as number of IBD targets) and we assume a detection efficiency of $100\%$. However, this channel is a big challenge; the main difficulty is that unlike IBD where a coincidence signature suppresses backgrounds, the proton scattering appears as a single flash of light with a much higher single-event backgrounds rate \cite{An:2015jdp}. The relevant backgrounds arise from a variety of radioactive decays as well as cosmogenic backgrounds and solar neutrinos. The DSNB background in the $\nu-p$ channel of JUNO has not been studied in depth yet, and it remains to be determined how well factors such as material distillation, fiducial cuts, and pulse shape discrimination will mitigate these backgrounds \cite{Li:discuss}. As an optimistic scenario, we therefore follow the Galactic search where scintillator detectors are expected to significantly suppress background once a cut of $T_{\rm ke}>0.2$ MeV on the proton kinetic energy is imposed \cite{Dasgupta:2011wg}. Hence, we assume this will be also the case at JUNO and calculate the events at the range $0.2-1.5$ MeV. We expect that JUNO will collect a total of  $53~(98)$ $\nu-p$ events using the H18 (thermal FD) model. We show in Fig. \ref{fig:JUNOnup} the distribution of events for total number of $\nu-p$ events. We will first show results without this JUNO channel, and later including this channel. 

\section{Results}\label{sec:results}
In this section we exploit the flavor composition of the DSNB flux defining the ratio of the $\nu_e$, $\bar\nu_e$ and $\nu_x$ fluxes to the total flux.  To see what is the potential of the DUNE, Hyper-K, and JUNO experiments to discover the DSNB flavor content, we perform a $\chi^2$ analysis considering the systematic uncertainties using a pull method. The goal is to see what is the flavor content using the MSW mechanism, considering the DSNB fluxes discussed in Section \ref{sec:flux}, and compare it with the case where neutrino decay has happened. To do this we define the flavor parameter $f_\alpha$ as the ratio of the fluxes of $\nu_e$, $\bar\nu_e$ and $\nu_x$ to the total flux \cite{Arguelles:2015dca}, that is,
\bea
f_\alpha\equiv\frac{\phi_\alpha}{\sum_{\beta=\nu_e,\bar{\nu}_e,\nu_x} \phi_\beta},
\eea
where $\phi_\alpha$ is the flux of $\nu_\alpha$ for either the MSW or the decay scenarios. We have checked that in the range of neutrino energies considered in our studies the energy dependence of $f_\alpha$ is less than $3\%$, so we assume it is a constant. We calculate the theoretical number of events at each energy bin $i$ as $N_i^{\rm{th}}(f_\alpha)$ using Eqs.~(\ref{eq:DUNEevents}), (\ref{eq:rateIBD}), (\ref{eq:HKnueevents}) and  (\ref{eq:JUNOnupevents}) but replacing the DSNB fluxes with $f_\alpha \sum_\beta \phi_\beta$. We define the $\chi^2$ function as
\begin{align}
  \chi^2 =\sum_i \frac{\Big((1+a)N_i^{\rm{th}}+(1+b) B_i-N_i^{\rm{DSNB}}-B_i\Big)^2}{B_i+N_i^{\rm{DSNB}}}+\frac{a^2}{\sigma_a^2}+\frac{b^2}{\sigma_b^2},
  \end{align}
where $N_i^{\rm{DSNB}}$ is the number of DSNB events at each bin $i$ that we calculate assuming that the true theory is MSW and then the decay scenario. The background at each bin is given by $B_i$ and is considered after all background reduction methods have been taken into account. The pull parameters $a$ and $b$ take into account the flux and background uncertainties, and we take their errors to be $\sigma_a=30\%$ and $\sigma_b=20\%$, respectively. The former is a reflection of the DSNB uncertainty, mainly driven by the uncertain core-collapse rate; while the latter is a reflection of the experimental backgrounds which depend on the detector and channel, and we assume 20\% in light of the uncertainty in the low-energy ($<1$ GeV) atmospheric neutrino flux \cite{Honda:1995hz} which contributes dominantly towards the background in most cases (including, e.g., invisible muons, atmospheric CC and NC).

\begin{figure}[h!]
\centering
\includegraphics[width=1\textwidth]{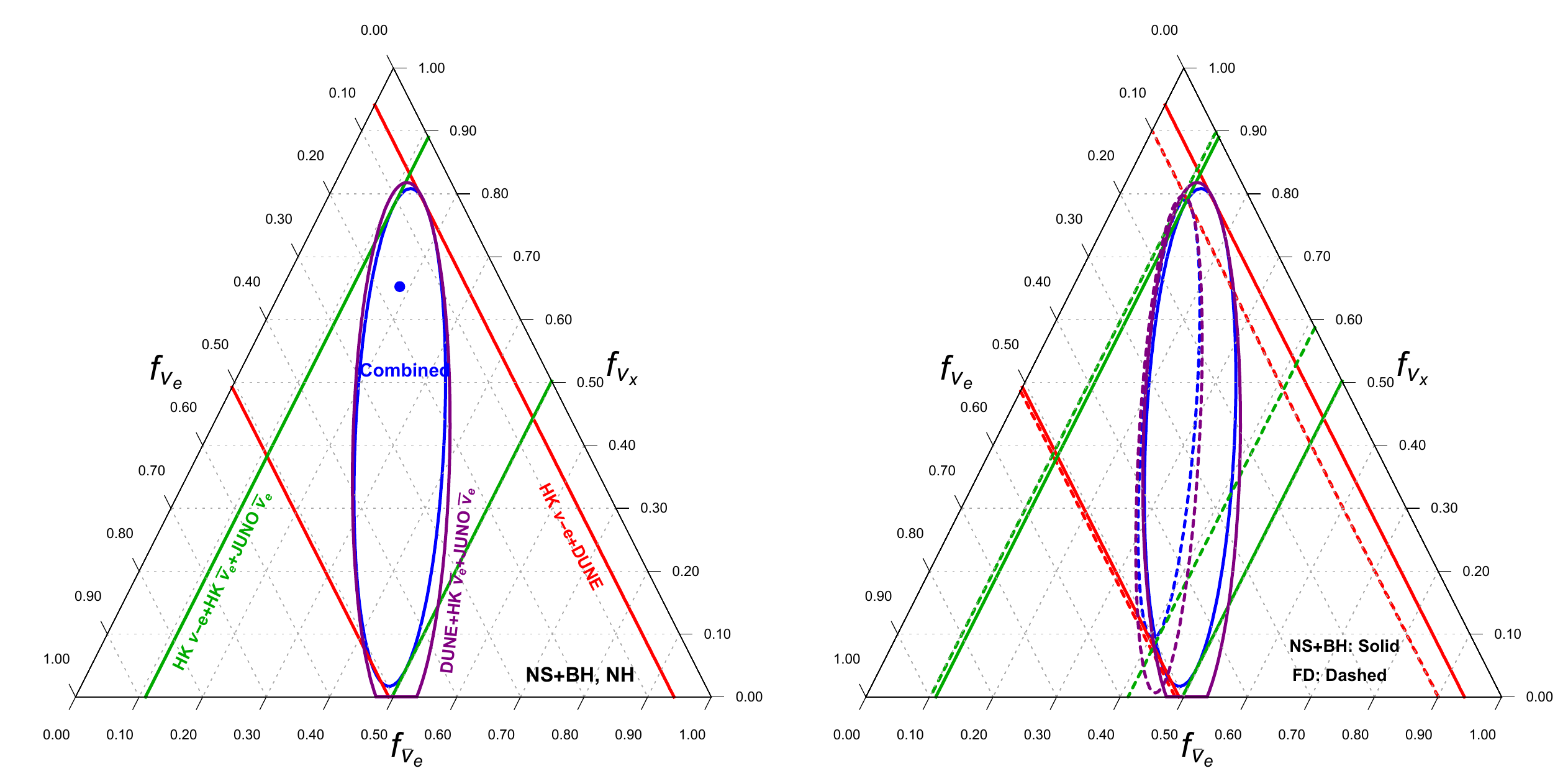}
\caption{\label{fig:triangle1}  Flavor composition of the DSNB neutrinos at earth for the MSW scenario in NH. All the curves are at $90\%$ C.L.. Left panel: The combined results of Hyper-K (IBD events plus $\nu-e$ events) and JUNO (IBD events) is shown in green, of Hyper-K ($\nu-e$ events) and DUNE (CC events) is shown in red, and Hyper-K (IBD events) and JUNO (IBD events) and DUNE (CC events) is shown in purple. The combined result of all four data sets are shown in blue. The blue dot shows the best fit point: $(f_{\nu_e},f_{\bar\nu_e},f_{\nu_x})=(0.17,0.18,0.65)$. Right panel: The comparison of the H18 simulation (solid) with the thermal FD distribution adopting $T_{\nu_e}=5$~MeV, $T_{\bar\nu_e}=6$~MeV and $T_{\nu_x}=7$~MeV (dashed). }
\end{figure}

We  first show  our  results for the progenitor-averaged DSNB model of H18 with MSW mixing in the left panel of Fig.~\ref{fig:triangle1}. These results are obtained without the $\nu-p$ events at JUNO. The full 3-detector 4-channel combination is shown by the blue contour, while subset combinations are shown and labeled in other colours. We find the best fit values of $(f_{\nu_e},f_{\bar\nu_e},f_{\nu_x})=(0.17,0.18,0.65)$. The bound on $f_{\nu_e}$ mainly comes from the DUNE experiment, while the IBD events at Hyper-K and JUNO give constraints on $f_{\bar\nu_e}$.  The $90\%$ C.L. constraints on these two parameters are 
\bea 
0.08<f_{\nu_e}<0.37,\quad 0.12<f_{\bar\nu_e}<0.36,
\eea
while the $f_x$ is not strongly constrained. This is because the constraint on $f_x$ comes from the $\nu-e$ events at Hyper-K, which has both low statistics and large backgrounds.  The right panel of Fig.~\ref{fig:triangle1} compares the MSW scenario for the NS$+$BH model of H18 with the thermal FD model. As expected, the latter gives stronger constraints on the parameters due to its larger event statistics. In this case the best fit values are $(f_{\nu_e},f_{\bar\nu_e},f_{\nu_x})=(0.19,0.16,0.65)$ and we find the following $90\%$ C.L. intervals:  
\bea 
0.11<f_{\nu_e}<0.39,\quad 0.10<f_{\bar\nu_e}<0.32.
\eea

\begin{figure}[h!]
\centering
\includegraphics[width=1\textwidth]{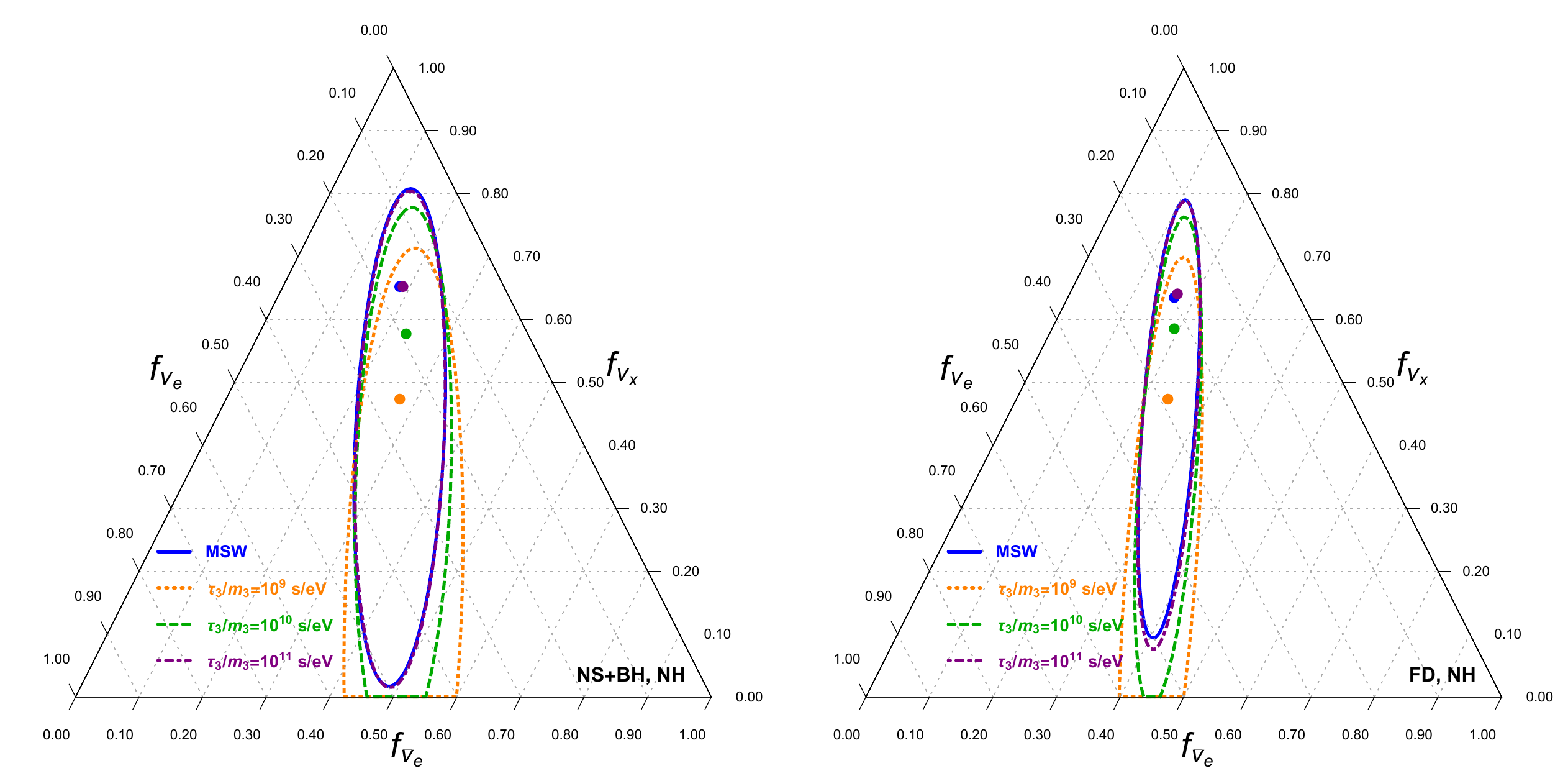}
\caption{\label{fig:triangledeccay1}  The $90\%$ C.L. flavor contours for the MSW mechanism (blue) and the neutrino decay scenario for three different values of $\tau_3/m_3$. The left and right panels denote to the progenitor averaged H18 and thermal FD fluxes, respectively.  }
\end{figure}

We show in Fig.~\ref{fig:triangledeccay1}
our obtained results for the neutrino decay for different values of $\tau_3/m_3$. As we discussed in the previous section, the faster neutrinos decay we expect higher number of $\nu_e$ and $\bar\nu_e$ events in terrestrial detectors. This translates into larger $f_{\nu_e}$ and $f_{\bar\nu_e}$, but smaller values for $f_{\nu_x}$. This can be seen by the orange curve in Fig.~\ref{fig:triangledeccay1} which corresponds to $\tau_3/m_3=10^9$~s/eV. The obtained best fit values for $(f_{\nu_e},f_{\bar\nu_e},f_{\nu_x})$ are:
\bea
\tau_3/m_3&=&10^9{\rm{s/eV}}:~~~(0.24,0.27,0.49),\nonumber\\
\tau_3/m_3&=&10^{10}{\rm{s/eV}}:~~(0.19,0.21,0.60),\\
\tau_3/m_3&=&10^{11}{\rm{s/eV}}:~~(0.17,0.18,0.65).
\eea

\begin{figure}[h!]
\centering
\includegraphics[width=1\textwidth]{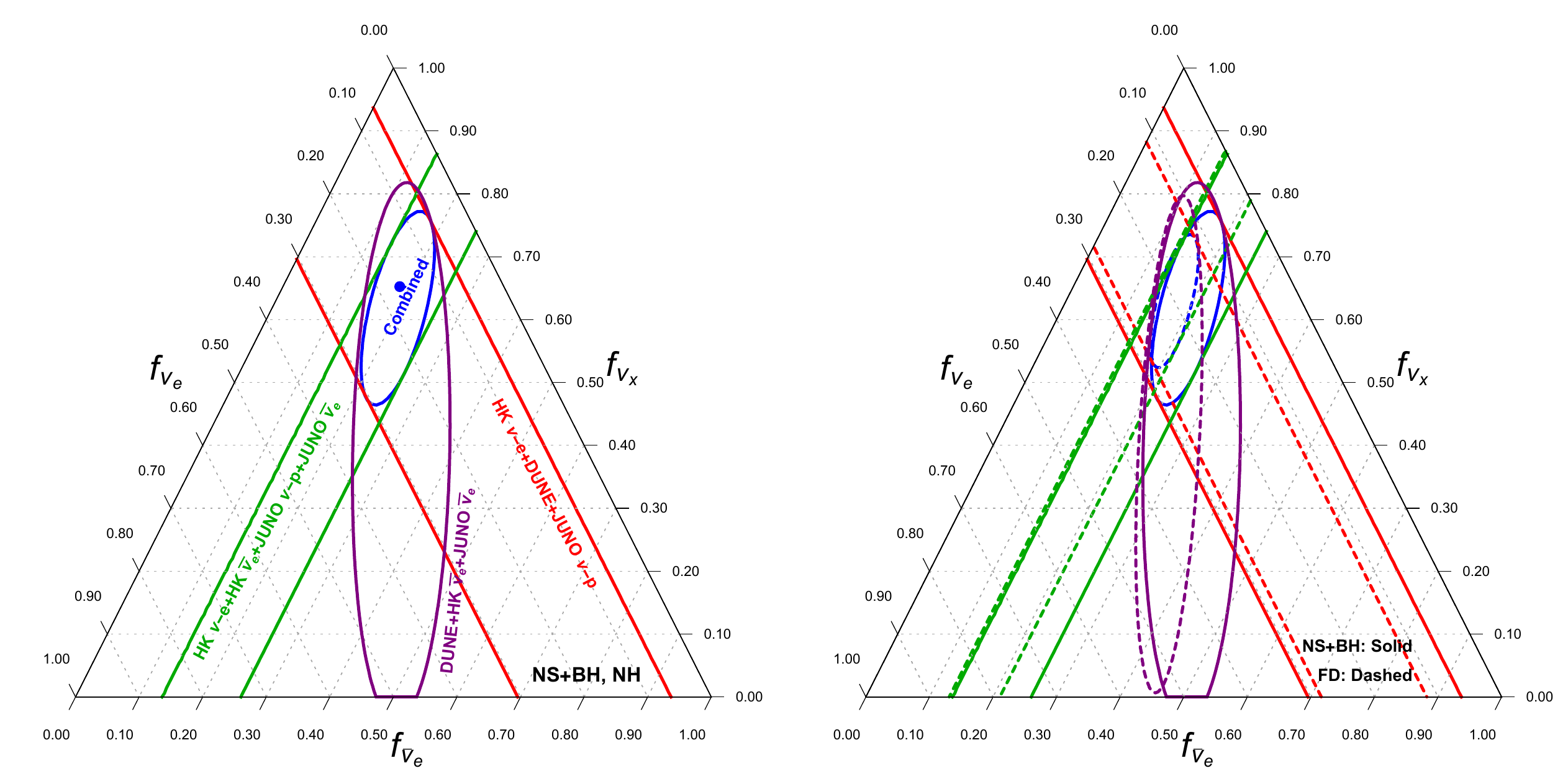}
\caption{\label{fig:triangle2}  Same as Fig.~\ref{fig:triangle1}, but the $\nu-p$ events at JUNO are also added. See text for details. }
\end{figure}

If the JUNO experiment can successfully mitigate its single-background rate for the $\nu-p$ events then we can get much stronger constraints on the flavor parameters, specifically on $f_{\nu_x}$ which is largely unconstrained otherwise. We show the results in Fig.~\ref{fig:triangle2}.  In this case we obtain the following $90\%$ C.L. constraints:
\bea 
0.08(0.11)<&f_{\nu_e}&<0.37(0.38),\nonumber\\ 0.11(0.10)<&f_{\bar\nu_e}&<0.34(0.29),\\
0.35(0.38)<&f_{\nu_x}&.\nonumber
\eea
for the progenitor averaged H18 (thermal FD) model. 

\begin{figure}[h!]
\centering
\includegraphics[width=1\textwidth]{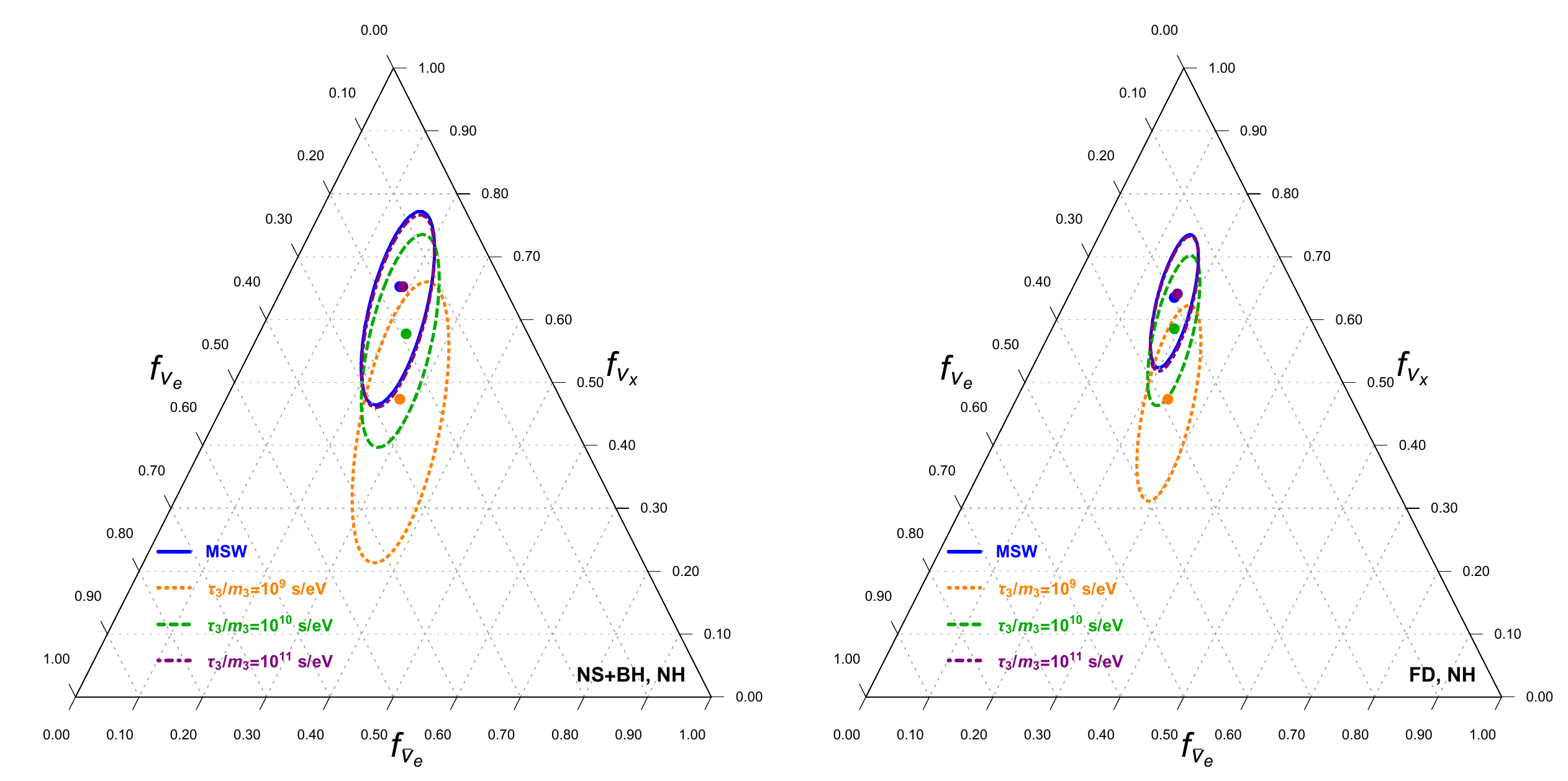}
\caption{\label{fig:triangledeccay2} Same as Fig.~\ref{fig:triangledeccay1} but including the $\nu-p$ events at JUNO. See text for details. }
\end{figure}

Finally, we show in Fig.~\ref{fig:triangledeccay2} our obtained results for the neutrino decay for different values of $\tau_3/m_3$, including $\nu-p$ events at JUNO. The higher statistic of all different flavors at JUNO helps in getting much better sensitivity to the decay parameters, in such a way that for $\tau_3/m_3=10^9$~s/eV, the corresponding allowed region has very little overlap with the MSW region, and can be excluded with $\sim90\%$ C.L..

\section{Conclusions}\label{sec:conslusion}
Although Galactic SNe are rare, happening only a few times per century, each hour a vast number of SN explosions happen over the whole universe, resulting in the DSNB. With the Gd enrichment at Super-K, the first statistically significant detection of the DSNB is on the horizon over the next decade, while next generation neutrino experiments which are already under construction, like Hyper-K, DUNE and JUNO, can observe an order of magnitude or more DSNB events thanks to their enormous detectors. Just as importantly, this opens multiple detection channels with sensitivity to multiple neutrino flavors.

In this paper we studied the flavor content of the DSNB in detail using the expected DSNB events at each of the aforementioned experiments. 
We considered two different neutrino emission models, the first the progenitor-averaged neutron star plus black hole (NS+BH) simulation of Ref.~\cite{Horiuchi:2017qja} (the H18 flux), and the second a simple Fermi-Dirac (FD) model described by a thermal temperature. The DSNB fluxes are compared in Fig.~\ref{fig:flux0}. 
If heavier neutrinos decay to lighter ones, we should expect a different DSNB flavor content arriving to the Earth from the standard MSW scenario. We considered such a decay scenario, described in detail in Sec.~\ref{sec:decay}, and we have compared the expected event rates to the MSW mixing without decay. 

The expected DSNB rates using our two DSNB models and two oscillation scenarios, for 20 years of data taking, are shown in Figs.~\ref{fig:DUNE}, \ref{fig:HK}, \ref{fig:HKnue}, \ref{fig:JUNOIBD} and \ref{fig:JUNOnup}. In the left panels, we show the results for only MSW, while in the right panels we show results for MSW plus neutrino decay for different values of $\tau_3/m_3$. We have also discussed the possible background sources for each of these experiments and the methods to suppress them. 

Based on our DSNB rate estimates for multiple neutrino experiments and detection channels, we study the flavor content of the DSNB using $f_\alpha$, the ratio of the flux of $\nu_\alpha$ over the total flux, at these experiments. The corresponding results are shown in Fig.~\ref{fig:triangle1}, where we quantified the $90\%$ C.L. allowed ranges for $f_\alpha$ comparing the progenitor-averaged NS+BH model of H18 and the thermal FD model. We find that a large fraction of the phase space will be tested out. In Fig.~\ref{fig:triangle2} we show the same, but this time including $\nu-p$ at JUNO. The $\nu-p$ at JUNO deserves a separate treatment, since it is unclear to what extent background can be mitigated, but is crucial for probing the $\nu_x$ content. 
Finally, we have shown the flavor triangles for the decay scenario in Figs.~\ref{fig:triangledeccay1} and \ref{fig:triangledeccay2}, with and without $\nu-p$ at JUNO, respectively. If the $\nu-p$ at JUNO can be realized, the corresponding allowed region for $\tau_3/m_3=10^9$~s/eV has little overlap with the no-decay MSW region, and can be excluded with $\sim90\%$ C.L.. 

The DSNB is a guaranteed flux of core-collapse neutrinos which is anticipated to be detected in the next several years by Super-K enhanced by Gd. The next decades will allow detailed studies with the DSNB made possible by future neutrino experiments such as the DUNE, Hyper-K, and JUNO. The flavor content of the DSNB will be a unique component of DSNB studies in this era. We have quantified how well the flavor content of the DSNB will be constrained by 20 years of running with future detectors. As theoretical uncertainties of the DSNB are reduced, the flavor information can become an important element to probe the physics of the DSNB.

\acknowledgments

We thank Huiling Li for fruitful discussions on JUNO backgrounds. We thank the anonymous referee for helpful comments. The work of Z.T.\ is supported by the U.S.\ Department of Energy under the award number DE-SC0020250 and DE-SC0020262. S.H.\ is supported by the U.S.\ Department of Energy Office of Science under award number DE-SC0020262 and NSF Grants Nos.\ AST-1908960 and PHY-1914409.

\bibliographystyle{apsrev4-1}
\bibliography{main}

\end{document}